\documentclass[aps,pre,twocolumn,superscriptaddress,showkeys,longbibliography,floatfix]{revtex4-2}

\usepackage[T1]{fontenc}
\usepackage{lmodern}
\usepackage[utf8]{inputenc}
\usepackage{amsmath,amssymb,bm,mathtools}
\usepackage{graphicx}
\usepackage{booktabs}
\usepackage{microtype}
\usepackage{xcolor}
\usepackage{hyperref}
\usepackage{siunitx}
\hypersetup{colorlinks=true,linkcolor=blue,citecolor=blue,urlcolor=blue}
\newcommand{\E}{\mathbb{E}}
\newcommand{\K}{\mathcal{K}}
\newcommand{\Lya}{\Lambda}

\newcommand{\Renv}{\mathrm{R}}
\newcommand{\Penv}{\mathrm{P}}
\begin{document}

\title{Shared environmental risk selects asymmetric inheritance of a protective reserve}
\author{Quentin Thommen}
\affiliation{CRCLille--Cancer Research Center of Lille, Universit\'e de Lille, UMR 9020 CNRS, Inserm U1366, CHU de Lille, France}
\date{\today}
\email{quentin.thommen@univ-lille.fr}

\begin{abstract}
  Environmental sharing changes the value of diversification even when the marginal statistics experienced by each lineage remain unchanged. A minimal model of cell division couples this effect to the inheritance of a conserved protective reserve. Each mother partitions its reserve between two daughters, and each fixed partition policy generates a random demographic operator whose top Lyapunov exponent determines long-term growth. Weak environmental sharing favors symmetric inheritance, whereas sufficiently shared innovation selects a separated asymmetric branch near $\alpha\simeq0.2$. The transition therefore occurs by a finite branch crossing rather than by a continuous departure from equal partition. The asymmetric phase persists when the protection law or reserve turnover is changed, although its boundary depends on protection nonlinearity and reserve memory. Because shared and private environmental innovations have identical marginal statistics, the mean reproductive operator is independent of the shared-innovation fraction. A dominant-mode second-order approximation then separates the asymmetric advantage into a mean-operator growth cost of specialization and a reduction of sensitivity to collective fluctuations. This approximation predicts the finite asymmetric branch selected by the full random-operator dynamics, while the full operator product determines the numerical crossing. Environmental sharing can therefore drive symmetry breaking in the inheritance of a conserved protective resource when the loss of diversification between lineages increases the value of diversification generated at division.
\end{abstract}

\maketitle

\section{Introduction}

Fluctuating environments select strategies that trade average performance against the risk of simultaneous demographic failure. In multiplicative population dynamics, long-term growth depends not only on mean reproductive output but also on temporal variance and correlations among the outcomes experienced by individuals or lineages \cite{Kussell2005,Starrfelt2012,Rivoire2011}. Phenotypic switching, plasticity, sensing, and memory can therefore increase long-term growth by distributing population states across uncertain futures \cite{Kussell2005,Rivoire2011,Belete2015,Mayer2017,Xue2018,George2026}. These theories usually treat diversification as a choice among phenotypic states or switching rules. The physical allocation of a conserved intracellular quantity introduces an additional constraint: increasing the amount inherited by one daughter necessarily decreases the amount inherited by the other.

Asymmetric segregation realizes this constraint directly. Unequal resource division reshapes stationary cellular distributions and population growth \cite{Marantan2016}, while asymmetric damage segregation can increase population fitness by repeatedly generating rejuvenated descendants \cite{Vedel2016,Pikovsky2023}. Protein-segregation models further show that nonlinear fitness effects can produce transitions between symmetric and asymmetric inheritance \cite{Lin2019}. Recent measurements in human colon cell lines also resolve cell-type-dependent cytoplasmic partition fluctuations and associate biased cytoplasmic inheritance with asymmetric daughter size, demonstrating that partition statistics can be quantified at population and single-cell scales \cite{Caudo2025}. Asymmetric partitioning is therefore experimentally accessible, and theory already predicts symmetry-breaking transitions under nonlinear fitness effects. These studies do not isolate environmental sharing, at fixed one-lineage statistics, as the control parameter selecting inheritance symmetry.

Environmental correlation changes the source of diversification available to a lineage. Bet-hedging theory distinguishes variation experienced independently by descendants from variation shared across a genotype, because correlated failures cannot be averaged across contemporaneous lineages \cite{Starrfelt2012}. This distinction does not require any change in the marginal frequency or persistence of favorable and adverse states. When daughter lineages experience largely independent futures, environmental heterogeneity itself diversifies their outcomes. When their futures become shared, this external diversification declines. Partition-generated heterogeneity can then increase long-term growth even though the marginal environment, reserve dynamics, and division costs remain unchanged.

\begin{figure}[t]
    \centering
    \includegraphics[width=\linewidth]{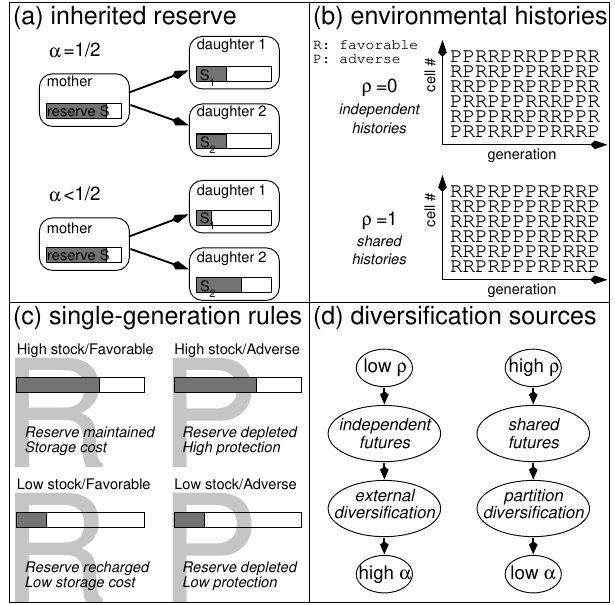}
    \caption{\textbf{Model ingredients and sources of diversification.}
    (a) A mother partitions a conserved protective reserve between two daughters. Symmetric inheritance, $\alpha=1/2$, gives equal daughter stocks, whereas $\alpha<1/2$ produces a reserve-poor and a reserve-rich daughter while conserving the transmitted material.
    (b) The shared-innovation fraction $\rho$ controls the similarity of the environmental histories subsequently experienced by different lineages. At $\rho=0$, environmental innovations are independent across lineages; at $\rho=1$, private innovations vanish and lineages in the same current environmental state share the same transition draw. $\Renv$ and $\Penv$ denote favorable and adverse states, respectively.
    (c) Reserve dynamics and reproduction depend jointly on inherited stock and current environment. Favorable conditions maintain or replenish reserve at a storage cost, whereas adverse conditions deplete reserve and convert high stock into protection.
    (d) Independent environmental histories provide external diversification between lineages. Shared histories reduce this source of diversification and increase the potential value of heterogeneity generated by reserve partition.}
    \label{fig:model_scheme}
\end{figure}

A conserved protective reserve turns this statistical distinction into an inheritance problem. The reserve carries a maintenance cost in favorable conditions but increases reproductive success under adverse conditions. Its partition obeys a conservation constraint, and recharge, depletion, and repeated inheritance propagate daughter differences across generations. Protective storage can itself provide an adaptive response to fluctuating metabolic demand \cite{Pfeuty2016}; division adds the requirement that stored material be allocated between descendants whose future environments may be correlated. Figure~\ref{fig:model_scheme} summarizes these ingredients. Equal partition produces similar daughter stocks, whereas asymmetric partition produces a reserve-poor and a reserve-rich daughter [Fig.~\ref{fig:model_scheme}(a)]. The shared-innovation fraction $\rho$ acts on subsequent environmental histories rather than on the partition event itself [Fig.~\ref{fig:model_scheme}(b)]. Reserve and environment then jointly control recharge, depletion, storage cost, and protection [Fig.~\ref{fig:model_scheme}(c)]. The resulting contrast is between diversification supplied by independent futures and diversification generated by partition when those futures become shared [Fig.~\ref{fig:model_scheme}(d)].

The model isolates this mechanism by varying only the fraction $\rho$ of environmental innovation shared across lineages. Each partition policy defines a positive random demographic operator acting on the joint distribution of environmental state and inherited reserve. The top Lyapunov exponent measures typical long-term population growth under the resulting sequence of shared environmental innovations. Direct evolutionary simulations first determine whether asymmetry emerges without prescribing candidate strategies. Operator calculations then resolve the competing growth branches, test their structural robustness, and identify the transition as a crossing between symmetric and finite-asymmetry demographic organizations.

The operator structure also separates the mechanism analytically. Because shared and private innovations obey identical marginal transition laws, the mean reproductive operator is independent of $\rho$; increasing the shared-innovation fraction changes the weight of the common fluctuating component without changing the mean operator. A dominant-mode second-order approximation therefore decomposes the asymmetric advantage into a mean-operator growth cost of specialization and a reduction of sensitivity to collective risk. This balance predicts the finite asymmetric branch even though redistribution through subdominant modes shifts the numerical crossing. The resulting framework links bet hedging under correlated risk to symmetry breaking in the inheritance of a conserved protective material.

\section{Model}
\subsection{State variables and sequence of events}
The model follows a large population structured by the environmental state experienced by each lineage and by the reserve inherited at birth. The density $d_t(E,S)$ denotes the population mass at generation $t$ in environmental state $E\in\{\Renv,\Penv\}$ with reserve $S\in[0,S_{\max}]$. State $\Renv$ is favorable and replenishes reserve, whereas state $\Penv$ is adverse and consumes it [Fig.~\ref{fig:model_scheme}(c)]. One generation follows a fixed sequence: the current reserve controls reproduction, the environment updates the reserve, the mother partitions the updated stock between two daughters, and each daughter receives the environment of the next generation. This ordering separates the reserve that protects the mother during the current episode from the reserve transmitted to the daughters.

\subsection{Reserve dynamics and environment-dependent reproduction}
The reference reserve map transforms the parental stock $S$ into the stock available at division,
\begin{align}
F_{\Renv}(S)&=\min\left[S_{\max},(1-\delta)S+R_0\right],\\
F_{\Penv}(S)&=\max\left[0,(1-\delta)S-D_0\right].
\label{eq:reserve_ref}
\end{align}
The leakage rate $\delta$ represents constitutive turnover. A favorable episode adds the increment $R_0$ until the capacity $S_{\max}$ is reached. An adverse episode removes the demand $D_0$ and clips the stock at zero. The ratios $R_0/H$ and $D_0/H$, where $H$ is the protective scale introduced below, determine whether protection can be built over several favorable generations and lost over one or several adverse generations.

The current reserve modifies reproduction in opposite directions in the two environments,
\begin{align}
w_{\Renv}(S)&=\exp(-c_S S),\\
w_{\Penv}(S)&=h(S).
\label{eq:fitness}
\end{align}
The favorable-state factor $\exp(-c_S S)$ assigns a continuous maintenance cost to stored material. The coefficient $c_S$ therefore controls the cost of carrying reserve even when no protection is required. The adverse-state function $h(S)$ converts reserve into survival or reproductive protection. The reference response is logistic,
\begin{equation}
h_{\mathrm{log}}(S)=h_-+(h_+-h_-)
\left[1+\exp\left(-\frac{S-H}{\varepsilon}\right)\right]^{-1},
\label{eq:logistic}
\end{equation}
where $H$ locates the protective transition and $\varepsilon$ sets its width. The lower and upper plateaus $h_-$ and $h_+$ fix the reproductive output of unprotected and protected cells. Hill, Michaelis--Menten, and linearly saturating responses test whether the transition requires a nearly discontinuous threshold.

\subsection{Hereditary partition and its costs}
A mother with post-environment stock $S'=F_E(S)$ generates two daughters with inherited reserves
\begin{equation}
S_1=\alpha S',\qquad S_2=(1-\alpha)S',\qquad 0\leq\alpha\leq\frac12.
\label{eq:partition}
\end{equation}
The restriction $\alpha\leq1/2$ only labels the poorer daughter as daughter 1; exchanging the daughters leaves the policy unchanged. Symmetric inheritance corresponds to $\alpha=1/2$. Decreasing $\alpha$ concentrates reserve in one daughter while depriving the other, but conserves the total transmitted stock $S_1+S_2=S'$. Heritability applies to the partition policy $\alpha$, whereas reserve-rich and reserve-poor daughter states are regenerated at each division.

Two distinct costs regulate reserve use. The storage factor
$\exp(-c_S S)$ penalizes the amount of material carried by each cell. It acts every favorable generation and therefore opposes policies that maintain a large stock over time. The polarization cost
\begin{equation}
C_A(\alpha)=c_A(1-2\alpha)^2
\label{eq:cost}
\end{equation}
assigns a phenomenological cost to unequal inheritance. It vanishes at symmetric division, increases quadratically with the difference between daughter fractions, and reaches $c_A$ at complete polarization. The reproductive contribution of a mother is multiplied by $\exp[-C_A(\alpha)]$. This cost does not remove reserve from the daughters; it reduces their total demographic production. The model therefore separates three effects that could otherwise be conflated: reserve conservation at division, maintenance of stored material through $c_S$, and the demographic cost assigned to unequal partition through $c_A$.

\subsection{Environmental correlation}
Each daughter receives a new environment through a mixture of global and private innovations. Let $P(E'|E)$ denote the prescribed marginal Markov transition. A Bernoulli selector chooses the global innovation with probability $\rho$ and a private innovation with probability $1-\rho$. For each current environmental state, all lineages share the corresponding global draw at generation $t$, whereas private draws average within the large population. The parameter $\rho$ is therefore the fraction of environmental innovation shared by lineages, not an imposed Pearson correlation coefficient. It controls cross-lineage correlation while preserving the transition probabilities experienced by each lineage separately.

The limits isolate two biological regimes. At $\rho=0$, future environmental innovations diversify across lineages and the population experiences the averaged transition. At $\rho=1$, environmental innovations are fully shared conditional on the current state, so contemporaneous lineages lose the diversification supplied by private draws. Intermediate values increase the shared component of environmental innovation, and hence cross-lineage covariance, without changing their one-lineage Markov statistics. The corresponding Pearson correlation is an emergent function of the Markov dynamics rather than the parameter $\rho$ itself.

\section{Random-operator formulation}
The operator formulation converts the biological rules above into the quantity selected by long-term population growth. Its main role is to preserve two facts simultaneously: private environmental variation averages over the large population, whereas the global innovation remains random from one generation to the next. A single deterministic transition matrix would erase the collective shocks that drive the proposed mechanism.

\subsection{One-generation demographic operator}
For a fixed global innovation $U_t$, partition policy $\alpha$, and shared-innovation fraction $\rho$, one generation applies a positive linear operator $\K_{U_t}^{(\alpha,\rho)}$ to the density $d_t$,
\begin{equation}
\widetilde d_{t+1}=\K_{U_t}^{(\alpha,\rho)}d_t.
\label{eq:operator}
\end{equation}
The operator executes, in order, environment-dependent reproduction, reserve update, partition into two daughter stocks, and transition to the daughters' environments. It therefore propagates the complete distribution over $(E,S)$ rather than only its mean. Different values of $\alpha$ generate different stock distributions and hence different demographic attractors.

The total population multiplication during generation $t$ is
\begin{equation}
G_t=\left\|\widetilde d_{t+1}\right\|_1,
\end{equation}
and the state used for the next iteration is normalized as
\begin{equation}
d_{t+1}=\frac{\widetilde d_{t+1}}{G_t}.
\label{eq:normalization}
\end{equation}
This normalization prevents numerical overflow while retaining the removed growth factor $G_t$ exactly.

\subsection{Typical long-term growth and policy selection}
The performance of policy $\alpha$ is the top Lyapunov exponent of the random operator product,
\begin{equation}
\Lya(\alpha,\rho)=\lim_{T\rightarrow\infty}
\frac{1}{T}\sum_{t=0}^{T-1}\log G_t.
\label{eq:lyapunov}
\end{equation}
This objective measures typical logarithmic growth along a realization of the shared environmental innovations. It differs from $\log\E[N_T]$, which can be dominated by rare favorable histories and can obscure the demographic consequence of collective adverse episodes.

The selected partition maximizes this typical growth rate,
\begin{equation}
\alpha^*(\rho)=\operatorname*{arg\,max}_{0\leq\alpha\leq1/2}
\Lya(\alpha,\rho).
\label{eq:optimum}
\end{equation}
The symmetric and asymmetric phases exchange order when
\begin{equation}
\max_{\alpha<1/2}\Lya(\alpha,\rho_c)
=\Lya(1/2,\rho_c).
\label{eq:frontier}
\end{equation}
Equation~\eqref{eq:frontier} is the central theoretical criterion. Each side must be evaluated on the distribution generated by its own policy; the transition cannot generally be inferred from a one-generation fitness comparison.

\section{Numerical strategy}
The stock interval is discretized on a uniform grid. Reserve updates and daughter stocks generally fall between nodes, so conservative linear deposition distributes each daughter contribution over its two neighboring bins. This procedure preserves nonnegativity and total mass before normalization. The same global environmental sequence is reused for all policies within a comparison; paired trajectories remove most of the sampling variance from differences in Lyapunov exponents.

Each run contains a burn-in phase that approaches the policy-specific random attractor and an accumulation phase that averages $\log G_t$. Independent global sequences provide standard errors. A discrete policy scan locates the competing maxima of $\Lya(\alpha,\rho)$, and local refinement resolves their crossing. Appendix~\ref{app:construction} gives the discrete operator and estimator, whereas Appendix~\ref{app:validation} reports parameters, convergence controls, and structural checks.

\section{Results}

\subsection{Collective risk selects a finite hereditary asymmetry}

Direct selection on the heritable partition trait tests whether asymmetric inheritance emerges without prescribing competing policies. Populations start with broad standing variation over $0\leq\alpha\leq1/2$, and weak mutation maintains local variation while demographic selection acts through the reserve dynamics, protection function, and polarization cost.

\begin{figure}[t]
\centering
\includegraphics[width=\linewidth]{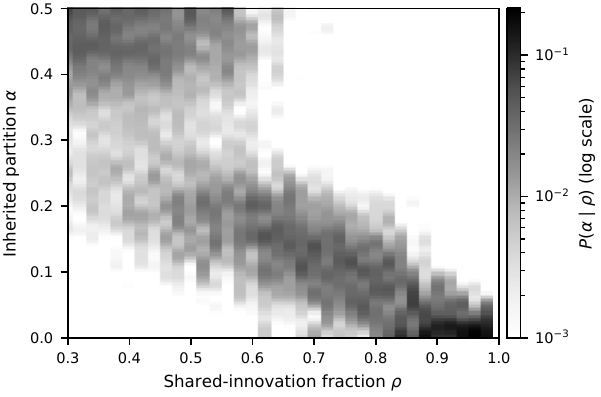}
\caption{Direct selection on the heritable reserve-partition trait. The grayscale map shows the terminal distribution $P(\alpha\mid\rho)$ obtained from polymorphic populations evolving under the reference reserve dynamics and protection function. Each column averages the last 200 generations and six independent replicates; the logarithmic grayscale retains low-frequency trait classes. Populations start with broad standing variation over $0\leq\alpha\leq1/2$, and weak mutation maintains local trait variation during selection. Weakly shared environmental innovation concentrates the distribution near symmetric inheritance, whereas increasing $\rho$ transfers probability to a separated asymmetric branch. Larger shared-innovation fractions subsequently shift this branch toward smaller $\alpha$, corresponding to stronger polarization of the maternal reserve.}
\label{fig:transition}
\end{figure}

Weakly shared environmental innovation concentrates the selected population near symmetric inheritance [Fig.~\ref{fig:transition}]. Increasing $\rho$ transfers probability to a separated branch near $\alpha\simeq0.2$, and stronger shared risk shifts this branch toward smaller $\alpha$. Selection therefore generates a distinct hereditary organization rather than a progressive broadening around $\alpha=1/2$. Mutation, finite trait sampling, and finite evolutionary time broaden this direct-selection transition, so Fig.~\ref{fig:transition} identifies the emerging branch but does not estimate $\rho_c$.

\begin{figure}[t]
\centering
\includegraphics[width=\linewidth]{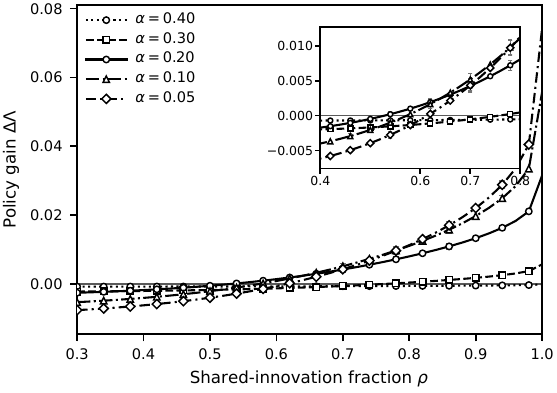}
\caption{Policy-specific long-term growth relative to symmetric inheritance. Each curve gives $\Delta\Lya(\alpha,\rho)=\Lya(\alpha,\rho)-\Lya(1/2,\rho)$ for a fixed partition policy under paired environmental sequences. Symmetry dominates when all curves remain below zero. As the shared-innovation fraction increases, a finite-asymmetry policy near $\alpha\simeq0.2$ becomes favorable while policies closer to $\alpha=1/2$ remain below the symmetric branch. More strongly polarized policies dominate only at larger $\rho$.}
\label{fig:frontier}
\end{figure}

Fixed-policy Lyapunov exponents identify the growth-rate structure underlying this transition. Figure~\ref{fig:frontier} plots
\begin{equation}
\Delta\Lya(\alpha,\rho)
=
\Lya(\alpha,\rho)-\Lya(1/2,\rho)
\label{eq:policy_gain}
\end{equation}
for several asymmetric policies.

At low $\rho$, every tested asymmetric policy remains below the symmetric reference [Fig.~\ref{fig:frontier}]. Near the transition, the $\alpha\simeq0.2$ branch crosses zero while policies closer to symmetry remain unfavorable. Larger shared-innovation fractions subsequently favor more polarized policies. The fixed-policy calculation therefore reproduces the strategy range selected in the polymorphic simulation and identifies the transition as a change in long-term growth ordering.

\subsection{Protection nonlinearity and reserve turnover delimit the asymmetric phase}

\begin{figure}[t]
\centering
\includegraphics[width=\linewidth]{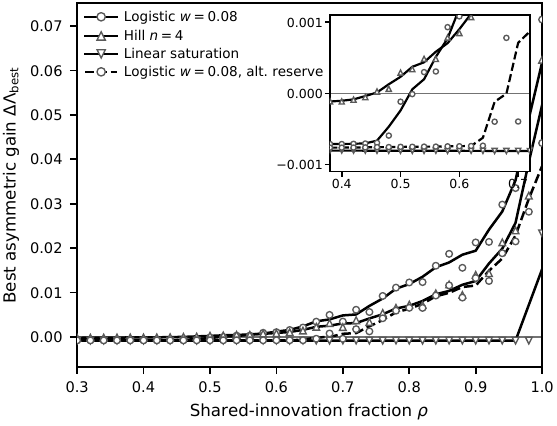}
\caption{\textbf{Robustness of the asymmetric advantage to protection shape and reserve dynamics.}
The gain $\Delta\Lya_{\rm best}$ of the best asymmetric policy over symmetric inheritance is shown for three protection laws under the reference reserve dynamics and for the reference logistic law under an alternative reserve map. The logistic response and Hill law with exponent four produce an interior transition, whereas linear saturation requires nearly complete shared risk before asymmetry becomes favorable. The dashed logistic curve uses capacity-limited recharge and multiplicative depletion and shows that altered reserve turnover shifts but preserves the transition. Symbols denote simulated mean gains; lines are three-point moving averages used only as guides to the eye. The inset resolves the sign changes around the transition region.}
\label{fig:reserve}
\end{figure}

The asymmetric phase is not restricted to the reference logistic law or to a single reserve-update rule. Figure~\ref{fig:reserve} compares the gain of the best asymmetric policy under three protection laws and one alternative reserve dynamics. The complete protection functions are shown in Appendix Fig.~\ref{fig:protection}.

Protection nonlinearity controls how efficiently reserve concentration separates daughter performance. Under the reference reserve dynamics, the narrow logistic response produces an interior transition, and a Hill response with exponent four retains the same qualitative behavior despite its different functional form [Fig.~\ref{fig:reserve}]. Linear saturation instead keeps the asymmetric gain near or below zero until collective risk is almost complete. The relevant requirement is therefore not a logistic threshold itself but a sufficiently nonlinear conversion of reserve into protection. This nonlinearity creates the potential benefit of reserve concentration but does not determine whether that benefit exceeds its cost; the shared-innovation fraction shifts this balance.

Reserve turnover provides an independent control of inherited memory. The alternative map
\begin{align}
F_{\Renv}^{(2)}(S)&=(1-\delta)S+R_0\left(1-\frac{S}{S_{\max}}\right),\\
F_{\Penv}^{(2)}(S)&=(1-\delta)S\exp(-D_0),
\label{eq:reserve_alt}
\end{align}
replaces constant recharge and additive depletion by capacity-limited recharge and multiplicative loss. The leakage rate remains $\delta$, whereas $D_0$ is reused here as a dimensionless depletion strength. Applied to the reference logistic protection law, this change shifts the zero crossing toward larger $\rho$ while preserving the increase of asymmetric advantage under strongly shared risk [Fig.~\ref{fig:reserve}]. Reserve persistence and protection nonlinearity therefore move the phase boundary while retaining the same qualitative increase of asymmetric advantage with environmental sharing.

\subsection{The transition is a crossing between separated growth branches}

\begin{figure}[t]
\centering
\includegraphics[width=\columnwidth]{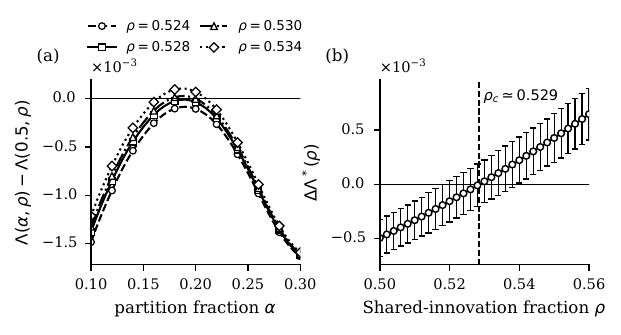}
\caption{\textbf{Crossing between symmetric and finite-asymmetry growth branches.}
(a) Growth-rate landscapes relative to symmetric inheritance for shared-innovation fractions surrounding the transition. The asymmetric maximum remains near $\alpha\simeq0.19$ while its height increases with $\rho$.
(b) Gain $\Delta\Lya^{*}(\rho)$ of the optimal asymmetric branch. The dashed line marks the zero crossing of the mean fine-scan estimate near $\rho\simeq0.53$; error bars are standard errors of paired differences across independent environmental sequences.}
\label{fig:scan_fine}
\end{figure}

The coarse policy comparison identifies a finite asymmetric branch but does not resolve its crossing with symmetry. A fine operator scan separates the branch position from its relative growth advantage. The asymmetric landscape retains a maximum near $\alpha\simeq0.19$ across the scanned correlations [Fig.~\ref{fig:scan_fine}(a)], while increasing $\rho$ primarily raises this maximum relative to the symmetric policy.

The optimal asymmetric gain,
\begin{equation}
\Delta\Lya^{*}(\rho)
=
\max_{\alpha<1/2}\Lya(\alpha,\rho)
-
\Lya(1/2,\rho),
\label{eq:best_asym_gain}
\end{equation}
changes sign when the two branches exchange order,
\begin{equation}
\Delta\Lya^{*}(\rho_c)=0.
\label{eq:rhoc_numeric}
\end{equation}
The simulations localize the crossing to $\rho\simeq0.52$--$0.53$, rather than resolving a universal or high-precision critical value. The mean gain changes sign near $\rho\simeq0.53$ in the fine scan [Fig.~\ref{fig:scan_fine}(b)]. Independent calculations with $N_S=121$ and $241$ preserve the finite optimum and place the sign change in the same narrow region, while stochastic sampling shifts the inferred zero within this interval. The robust result is therefore the crossing of a separated asymmetric branch, not the precise numerical value of $\rho_c$.

The separated maxima support a branch-crossing transition rather than a continuous drift away from equal inheritance. Increasing environmental sharing instead reverses the ordering of two demographic organizations generated by distinct operator products.

\subsection{A dominant-mode approximation separates the cost--buffering balance}
\label{sec:analytic_transition}

Environmental sharing enters the reproductive operator without changing its mean. For a fixed policy $\alpha$,
\begin{equation}
K_{\alpha,\rho}(u)
=
\overline K_{\alpha}
+
\rho Q_{\alpha}(u),
\qquad
\langle Q_{\alpha}(u)\rangle_u=0,
\label{eq:operator_decomposition}
\end{equation}
because global and private innovations obey the same marginal transition law. The mean operator $\overline K_{\alpha}$ is therefore independent of $\rho$.

Let $\lambda_{\alpha}$ denote the dominant eigenvalue of $\overline K_{\alpha}$ and $r_{\alpha}$ and $l_{\alpha}$ its right and left Perron eigenvectors, normalized by $l_{\alpha}^{T}r_{\alpha}=1$. Retaining only the component of the common fluctuation projected onto this Perron mode gives
\begin{equation}
\xi_{\alpha}(u)=l_{\alpha}^{T}Q_{\alpha}(u)r_{\alpha}.
\end{equation}
Expanding the resulting scalar growth factor to second order gives
\begin{equation}
\Lya_{\alpha}(\rho)
\simeq
\log\lambda_{\alpha}
-
\frac{\rho^{2}}{2}\sigma_{\alpha}^{2},
\qquad
\sigma_{\alpha}^{2}
=
\frac{\langle\xi_{\alpha}(u)^2\rangle_u}{\lambda_{\alpha}^{2}}.
\label{eq:lyapunov_second_order}
\end{equation}

Relative to symmetry,
\begin{equation}
\Delta\Lya_{\alpha}(\rho)
\simeq
-\Delta_0(\alpha)
+
\frac{\rho^2}{2}B(\alpha),
\label{eq:analytic_gain}
\end{equation}
with
\begin{equation}
\Delta_0(\alpha)
=
\log\frac{\lambda_{1/2}}{\lambda_{\alpha}},
\qquad
B(\alpha)
=
\sigma_{1/2}^{2}-\sigma_{\alpha}^{2}.
\label{eq:cost_buffering}
\end{equation}
The term $\Delta_0$ measures the mean-operator growth cost of asymmetric specialization, whereas $B$ measures the reduction of Perron-mode sensitivity to collective fluctuations. An asymmetric policy becomes competitive within this approximation only if the buffering gain compensates the mean-growth penalty. The approximation predicts
\begin{equation}
\rho_c^{(2)}(\alpha)
=
\left[
\frac{2\Delta_0(\alpha)}{B(\alpha)}
\right]^{1/2},
\qquad B(\alpha)>0,
\label{eq:rhoc_analytic}
\end{equation}
and selects the first competitive branch through
\begin{equation}
\alpha_c^{(2)}
=
\operatorname*{arg\,min}_{\alpha<1/2}
\rho_c^{(2)}(\alpha).
\label{eq:alphac_analytic}
\end{equation}

The approximation is used here to predict which asymmetric organization first becomes competitive, rather than to provide a quantitative estimate of the full random-product threshold. For the reference parameters, it gives $\alpha_c^{(2)}\simeq0.185$, matching the finite branch $\alpha^*\simeq0.18$--$0.19$ obtained from the full random-operator calculation. It predicts $\rho_c^{(2)}\simeq0.72$, above the numerical crossing near $0.52$--$0.53$. The dominant-mode approximation therefore captures the strategy selected by the cost--buffering balance but requires a larger shared-innovation fraction than the full operator product to reach the crossing. This discrepancy is consistent with quantitative contributions from multigenerational redistribution outside the dominant Perron mode, which the scalar approximation neglects.

\section{Discussion}

Environmental sharing selects between two distinct organizations of the same conserved reserve. Under weakly shared risk, symmetric inheritance maximizes growth while daughter lineages already diversify through their environmental histories. Increasing $\rho$ transfers a larger fraction of environmental innovation from private to shared fluctuations without changing the one-lineage transition statistics. This redistribution removes diversification between contemporaneous descendants and changes the value of heterogeneity generated at division. Asymmetric inheritance then becomes favorable because reserve concentration generates internal heterogeneity at division. The transition therefore links a statistical property of future environments to a physical symmetry of inheritance.

This mechanism extends classical bet-hedging arguments in a constrained setting. Bet-hedging theory establishes that long-term growth depends on correlations among individual outcomes because shared failures cannot be averaged across a population \cite{Starrfelt2012,Rivoire2011}. Phenotypic switching and plasticity exploit this principle by distributing descendants among alternative states \cite{Kussell2005,Belete2015,Mayer2017,Xue2018}. Reserve partition differs because the two daughter states cannot be chosen independently: conservation imposes $S_1+S_2=S'$. Increasing protection in one daughter necessarily decreases protection in the other. The model therefore converts the general value of diversification under correlated risk into a mechanically coupled allocation problem.

Asymmetric segregation itself is not specific to this mechanism. Unequal division can broaden resource distributions \cite{Marantan2016}, and damage segregation can increase population fitness by repeatedly generating rejuvenated descendants \cite{Vedel2016,Pikovsky2023}. Lin, Min, and Amir further showed that optimal protein segregation can undergo symmetry-breaking transitions when the segregated material changes cellular fitness nonlinearly \cite{Lin2019}. Measurements in colon cell lines now show that cytoplasmic partition fluctuations can differ strongly among cell types and that biased cytoplasmic inheritance can accompany asymmetric daughter size \cite{Caudo2025}. The present transition differs in its control parameter and in the role of the segregated material. The reserve is beneficial under adverse conditions and costly to maintain, whereas $\rho$ changes only the correlation of future environmental innovations. The switch from symmetric to asymmetric inheritance therefore occurs without changing the mean stress level, its persistence, or the total reserve transmitted at division. Within this controlled comparison, varying environmental sharing alone changes which inheritance organization maximizes long-term growth.

The finite position of the asymmetric branch follows from the competition between mean growth and collective-risk buffering. The polarization cost contributes to the position of the finite asymmetric optimum but cannot by itself generate its $\rho$-dependent reversal relative to symmetry. Its contribution depends on $\alpha$ but not on $\rho$; varying the shared-innovation fraction therefore changes the ordering of policies without changing their polarization penalty. The decomposition Eq.~\ref{eq:operator_decomposition} keeps $\overline K_\alpha$ independent of $\rho$ and assigns the effect of correlation to the fluctuating component. The dominant-mode approximation Eq.~\ref{eq:analytic_gain}
then separates the mean-growth penalty of specialization from its reduction of sensitivity to shared fluctuations. The ratio $\Delta_0(\alpha)/B(\alpha)$ selects a finite optimum because the policy that buffers collective risk most efficiently need not lie close to $\alpha=1/2$. Its minimum near $\alpha\simeq0.185$ reproduces the asymmetric branch of the full operator calculation. The same approximation overestimates the numerical crossing in $\rho$, indicating that dynamics outside the dominant-mode projection contribute substantially to the full threshold. Full-state redistribution, including reserve-memory effects, therefore modifies the strength of collective-risk buffering while leaving the selected finite branch close to the dominant-mode prediction for the reference parameters.

The structural controls delimit this mechanism rather than merely testing numerical robustness. A Hill protection law preserves the interior asymmetric phase, showing that the transition does not require the reference logistic form. Linear saturation strongly suppresses the gain from reserve concentration, while altered reserve turnover shifts the transition by changing how long daughter differences persist. Protection nonlinearity and reserve memory therefore determine how efficiently partition-generated heterogeneity survives and affects future reproduction. These dependencies make $\rho_c$ model dependent while preserving the broader prediction that shared risk can reorganize material inheritance.

The model deliberately excludes several processes that could modify the phase structure. Private environmental innovations are treated in the large-population limit, which neglects residual correlations generated by finite lineage clusters. Finite populations would additionally couple correlated environmental shocks to demographic extinction risk. Frequency-dependent ecological feedback could permit coexistence between partition strategies instead of selecting the single largest Lyapunov exponent. A stock-dependent policy could use maternal reserve as an informational variable and transform constant asymmetry into conditional allocation. Spatial structure would replace the abstract parameter $\rho$ by a relation among environmental correlation length, dispersal, and lineage separation. These extensions would change how shared risk is generated or perceived while retaining the conceptual distinction between diversification supplied by the environment and diversification generated at division.

The mechanism yields qualitative experimental signatures. Increasing the sharing of stress across related lineages while holding each lineage's marginal transition statistics fixed should favor stronger asymmetry in inherited protective material. Reducing reserve lifetime should weaken this response, whereas a sharper dependence of protection on reserve should strengthen it. The relevant observable is the daughter-to-daughter distribution of inherited material rather than its population mean. Recent flow-cytometry and time-lapse measurements demonstrate that such partition statistics can be resolved at population and single-cell scales \cite{Caudo2025}. A system in which partition asymmetry increases with environmental sharing at fixed marginal stress statistics would therefore provide a direct test of the proposed mechanism.

\section{Conclusion}

Environmental sharing can change the optimal symmetry of material inheritance without changing the marginal environment experienced by any lineage. A conserved protective reserve couples daughter phenotypes mechanically: equal partition exploits diversification supplied by independent futures, whereas unequal partition generates diversification when those futures become shared. Random demographic operators resolve this change as a crossing between symmetric and finite-asymmetry growth branches, while a dominant-mode approximation isolates the balance between specialization cost and collective-risk buffering. The mechanism therefore predicts that the inheritance pattern of protective material should respond to the degree to which future stress is shared across lineages, not only to its average intensity.

\appendix
\section{Operator construction and numerical controls}
\label{app:construction}

\subsection{Discrete state space and conservative deposition}
Let $S_i=i\Delta S$, $i=0,\ldots,N_S-1$, with $\Delta S=S_{\max}/(N_S-1)$. The state vector contains the masses $d_{E,i}$. For each source state $(E,i)$, reproduction contributes
\begin{equation}
q_{E,i}=2w_E(S_i)\exp[-C_A(\alpha)].
\end{equation}
The post-environment stock is $S'=F_E(S_i)$ and the daughter stocks are $x_1=\alpha S'$ and $x_2=(1-\alpha)S'$. When $x$ falls between two stock nodes, conservative linear interpolation deposits the corresponding mass on the neighboring bins. This rule preserves nonnegativity and total daughter mass.

\subsection{Global and private environmental transitions}
Conditional on the current state $E$, the effective transition during generation $t$ is
\begin{equation}
T_{U_t}^{(\rho)}(E'|E)=
(1-\rho)P(E'|E)+\rho\,\mathbf{1}_{E'=U_t(E)}.
\label{eq:transition_decomp}
\end{equation}
The global draw $U_t$ follows the same marginal transition law $P(E'|E)$ as the private component. Varying $\rho$ therefore preserves the one-lineage Markov statistics while changing the fraction of environmental innovation shared across the population. The resulting Pearson correlation between lineage states is an emergent function of $\rho$, the current environmental state, and the Markov transition probabilities; $\rho$ itself is the model control parameter.

\subsection{Lyapunov estimator and paired comparisons}
Starting from a normalized nonnegative state,
\begin{align}
\widetilde d_{t+1}&=\K_{U_t}^{(\alpha,\rho)}d_t,\\
G_t&=\sum_{E,i}\widetilde d_{t+1}(E,i),\\
d_{t+1}&=\widetilde d_{t+1}/G_t.
\end{align}
After burn-in, the estimator is
\begin{equation}
\widehat\Lya=\frac{1}{T}\sum_t\log G_t.
\end{equation}
The same global sequence is reused for all policies in a comparison. Paired differences
\begin{equation}
\Delta\widehat\Lya_r(\alpha)=
\widehat\Lya_r(\alpha)-\widehat\Lya_r(1/2)
\end{equation}
therefore remove most environmental sampling noise.

\subsection{Numerical protocols}
Table~\ref{tab:numerics} records the simulation lengths, burn-in periods, stock-grid resolutions, and replicate numbers used for each calculation. The direct polymorphic simulation uses $6000$ cells, $1200$ generations, a $300$-generation burn-in, six replicates, mutation probability $0.005$, and mutational standard deviation $0.003$ in $\alpha$. Its terminal distributions average the last $200$ generations. Fixed-policy scans use the discrete operator rather than finite populations.

\begin{table*}[t]
\caption{Numerical protocols used in the main calculations. The polymorphic calculation is individual based and therefore has no stock-grid resolution $N_S$.}
\label{tab:numerics}
\begin{ruledtabular}
\begin{tabular}{lccccc}
Calculation & $N_S$ & Generations & Burn-in & Replicates & Policy/risk grid\\
\colrule
Fig.~\ref{fig:transition}: polymorphic selection & -- & $1200$ & $300$ & $6$ & $\rho=0.30:0.02:1.00$\\
Fig.~\ref{fig:frontier}: fixed policies & $81$ & $1100$ & $200$ & $6$ & $\rho=0.30:0.02:1.00$\\
Fig.~\ref{fig:reserve}: structural controls & $81$ & $1400$ & $250$ & $6$ & $\rho=0.30:0.02:1.00$\\
Fig.~\ref{fig:scan_fine}: fine crossing scan & $121$ & $4000$ & $800$ & $8$ & $\rho=0.500:0.002:0.560$\\
Grid-convergence control & $81,121,241$ & $4000$ & $800$ & $8$ & $\rho=0.520,0.528,0.530,0.540$\\
Dominant-mode approximation & $121$ & -- & -- & -- & $\alpha=0.05:0.005:0.50$\\
\end{tabular}
\end{ruledtabular}
\end{table*}

\subsection{Convergence near the branch crossing}
The fine scan uses $N_S=121$ stock bins. A targeted grid control repeated the calculation at $N_S=81$, $121$, and $241$ for $\rho=0.520$, $0.528$, $0.530$, and $0.540$, with the asymmetric maximum refined over $0.17\leq\alpha\leq0.21$. The $N_S=121$ and $241$ calculations give nearly identical optimal policies and gains throughout this interval. Their inferred zero crossings fall near $\rho\simeq0.52$, whereas an independent fine scan places the mean crossing near $\rho\simeq0.53$. This variation is consistent with sampling uncertainty between nearly equal Lyapunov exponents at the crossing rather than with stock discretization. The finite asymmetric branch and the sign reversal away from the immediate zero are stable for $N_S\geq121$.

\section{Reference parameters and structural controls}
\label{app:validation}

Table~\ref{tab:params} lists the reference parameters. The protective threshold $H=1$ fixes the reserve unit. The reserve scale enters through dimensionless combinations such as $R_0/H$, $D_0/H$, $S_{\max}/H$, $\varepsilon/H$, and $c_S H$.

\begin{table}[h]
\caption{Reference model and numerical parameters.}
\label{tab:params}
\begin{ruledtabular}
\begin{tabular}{lcc}
Parameter & Symbol/code name & Value\\
\colrule
Maximum reserve & $S_{\max}$ & $4$\\
Recharge increment & $R_0$ & $0.45$\\
Adverse demand & $D_0$ & $0.80$\\
Leakage rate & $\delta$ & $0.02$\\
Protection threshold & $H$ & $1$\\
Logistic width & $\varepsilon$ & $0.08$\\
Low protection & $h_-$ & $0.08$\\
High protection & $h_+$ & $0.95$\\
Storage cost & $c_S$ & $0.025$\\
Polarization cost & $c_A$ & $0.02$\\
Favorable persistence & $q_R$ & $0.85$\\
Adverse persistence & $q_P$ & $0.75$\\
Reference stock bins & $N_S$ & $121$\\
\end{tabular}
\end{ruledtabular}
\end{table}

A simultaneous rescaling of $S_{\max}$, $R_0$, $D_0$, $H$, and $\varepsilon$, together with the inverse rescaling of $c_S$, leaves these dimensionless combinations unchanged; the absolute reserve unit therefore carries no independent information. Recharge, depletion, polarization cost, and environmental persistence alter reserve memory or the cost of maintaining unequal descendants, so $\rho_c$ is a model-dependent threshold rather than a universal constant.

\section{Protection functions used in the structural tests}

\begin{figure}[t]
\centering
\includegraphics[width=\linewidth]{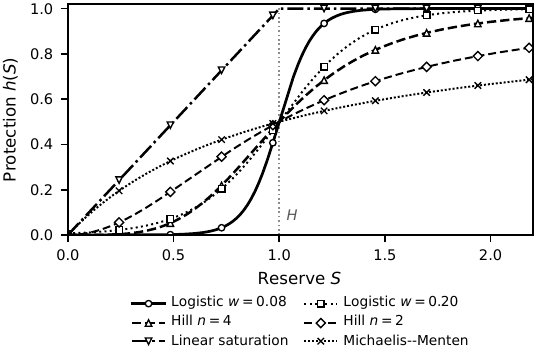}
\caption{Protection functions used in the structural tests. The vertical scale gives adverse-state protection $h(S)$ and the horizontal scale gives inherited reserve. The main text compares the reference logistic law, Hill $n=4$, and linear saturation; the additional curves document broader structural controls used to separate threshold sharpness from general nonlinear protection.}
\label{fig:protection}
\end{figure}

The structural tests probe which properties of the reserve--protection map
support the asymmetric advantage induced by environmental sharing.
 Figure~\ref{fig:protection} compares the protection functions used in these tests and separates the effect of threshold shape, steepness, and saturation from the reference logistic response. This comparison defines the functional perturbations underlying the robustness analysis discussed in the main text.

\end{document}